\def \SAIT #1 #2 {{\em Mem.\ Soc.\ Astron.\ It.\/} {\bf #1}, #2}
\def \MESS #1 #2 {{\em The Messenger\/} {\bf #1}, #2}
\def \ASTRNACH #1 #2 {{\em Astron. Nach.\/} {\bf #1}, #2}
\def \AAP #1 #2 {{\em Astron. Astrophys.\/} {\bf #1}, #2}
\def \AAL #1 #2 {{\em Astron. Astrophys. Lett.\/} {\bf #1}, L#2}
\def \AAR #1 #2 {{\em Astron. Astrophys. Rev.\/} {\bf #1}, #2}
\def \AAS #1 #2 {{\em Astron. Astrophys. Suppl. Ser.\/} {\bf #1}, #2}
\def \AJ #1 #2 {{\em Astron. J.\/} {\bf #1}, #2}
\def \ANNREV #1 #2 {{\em Ann. Rev. Astron. Astrophys.\/} {\bf #1}, #2}
\def \APJ #1 #2 {{\em Astrophys. J.\/} {\bf #1}, #2}
\def \APJL #1 #2 {{\em Astrophys.. J. Lett.\/} {\bf #1}, L#2}
\def \APJS #1 #2 {{\em Astrophys. J. Suppl.\/} {\bf #1}, #2}
\def \APSS #1 #2 {{\em Astrophys. Space Sci.\/} {\bf #1}, #2}
\def \ASR #1 #2 {{\em Adv. Space Res.\/} {\bf #1}, #2}
\def \BAIC #1 #2 {{\em Bull. Astron. Inst. Czechosl.\/} {\bf #1}, #2}
\def \JSQRT #1 #2 {{\em J. Quant. Spectrosc. Radiat. Transfer\/} {\bf #1}, #2}
\def \MN #1 #2 {{\em Mon. Not. R. Astr. Soc.\/} {\bf #1}, #2}
\def \MEM #1 #2 {{\em Mem. R. Astr. Soc.\/} {\bf #1}, #2}
\def \PLR #1 #2 {{\em Phys. Lett. Rev.\/} {\bf #1}, #2}
\def \PASJ #1 #2 {{\em Publ. Astron. Soc. Japan\/} {\bf #1}, #2}
\def \PASP #1 #2 {{\em Publ. Astr. Soc. Pacific\/} {\bf #1}, #2}
\def \NAT #1 #2 {{\em Nature\/} {\bf #1}, #2}
\documentstyle{memsait}
\input epsf.sty
\begin{opening}
\title{REM Optical Slitless Spectrograph (ROSS): an instrument for
prompt low resolution spectroscopy of Gamma Ray Bursts}
\author{Eliana Palazzi$^1$ and Elena Pian$^{1,2}$\\ on behalf of the ROSS 
team}
\institute{$^1$ Istituto TeSRE, CNR, via Gobetti 101, I-40129 Bologna, 
Italy\\
$^2$ Osservatorio Astronomico di Trieste, via G.B.
Tiepolo 11, I-34131 Trieste, Italy}
\date{} % DO NOT INSERT ANY DATE HERE !!!
\end{opening}

\begin{document}

%\oddpagefooter{\sf Mem. S.A.It., Vol. ??, ??}{}{\thepage}
%\evenpagefooter{\thepage}{}{\sf Mem. S.A.It., Vol. ??, ??}
\oddpagefooter{}{}{} % LEAVE AS IT IS !
\evenpagefooter{}{}{} % LEAVE AS IT IS !
\ 
\bigskip

\begin{abstract}
Since the discovery of Gamma-Ray Bursts (GRBs), attempts have been made to
detect correlated optical transient emission from these objects.  In January
1999, the ROTSE I robotic telescope detected a bright optical flash
simultaneous with a GRB, thanks to the prompt dissemination to
the ground of the high energy
event coordinates. To date, that single observation remains unique as no
other prompt flashes have been seen for other bursts observed with
comparably short response times. This suggests that in general GRB prompt
optical emission may be considerably dimmer than observed for the GRB990123
event. To exploit the better angular localization accuracy of the flying
(HETE-2) or soon to fly (INTEGRAL, AGILE, SWIFT, GLAST) missions for high
energy astrophysics, a new generation of robotic telescopes is being
developed.  These will have response times as short as a few seconds and
will be sensitive to signals as faint as $m_v\sim$ 20, thus increasing the
chance of detecting even weak prompt emission. Results from these
experiments should provide important new data about the dynamics and local
medium composition of GRBs.  In this paper we describe one of the new
instruments, ROSS, to be mounted on the robotic facility REM, designed to
perform low resolution spectroscopy on the prompt optical flares associated
with GRBs.
\end{abstract}

\section{Introduction}

Gamma Ray Bursts (GRBs), brief and intense flashes of gamma-rays
coming from random directions in the sky, were discovered
accidentally
more than thirty years ago. They are among the most mysterious
astronomical phenomena ever observed and only during the last decade we
started understanding their physics. First, the BATSE instrument on the
Compton Gamma
Ray Observatory (1991-2000) discovered that GRBs are distributed
isotropically over the sky, thus suggesting their extragalactic origin
[8]. Starting in 1997, the combined capabilities of
the Gamma Ray Burst Monitor (GRBM)  and Wide Field Cameras (WFCs)
onboard BeppoSAX enabled unprecedentedly accurate angular localization
of GRBs, thus allowing longer wavelengths (radio, optical and X-ray)
follow-up observations (e.g. [5,9,23]). These observations led to the
discovery of delayed, fading
emission (the afterglow) and eventually confirmed, 
via 
redshifts measurements [17], that GRBs are located at 
cosmological distances.  This implies that huge 
energies 
($10^{51}-10^{54}$ ergs, assuming isotropic emission)
are released during the gamma-ray event and the multiwavelength afterglow,
supporting the relativistic
fireball scenario [4,18,19].  However, the physics of GRBs and of their
lower energy counterparts presents still many obscure aspects.
\begin{figure}
\epsfysize=7cm % fix the y-dimension and scales x-dim. to y-dim.
%\epsfxsize=8cm % fix the x-dimension and scales y-dim. to x-dim.
% Feel free to do the choice you prefer but do not exceed the x-dimension
% of the text lines
\hspace{2.cm}\epsfbox{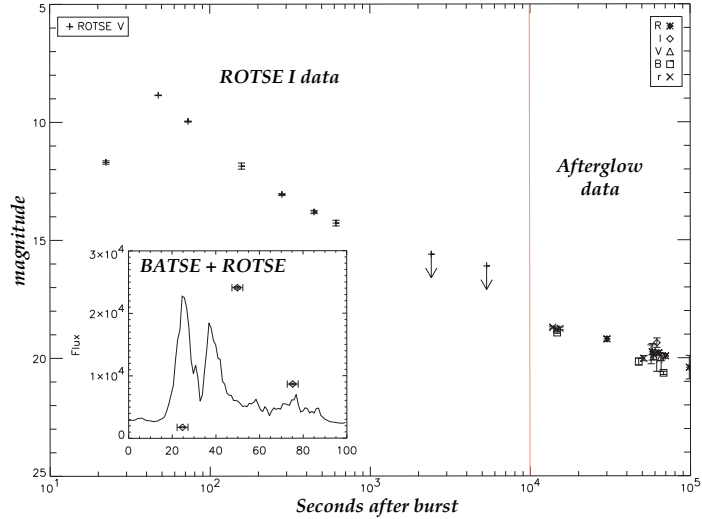} 
%for centering: act on hspace argument 
\vspace{-0.3cm}
\caption[h]{Light curve from ROTSE I observations and some early
afterglow points. The insert shows the first three ROTSE points
superimposed to the BATSE gamma-ray light curve. The vertical line
divides the ROTSE I from the afterglow observations and approximately
indicates when the
earliest arcminute position was made available by BeppoSAX (from [12]).}
\end{figure}

\section{Fast follow-up of optical/near-infrared flashes of Gamma-Ray
Bursts: bridging the gap between the explosion and its afterglow}

According to the general interpretation,
the prompt high energy emission of GRBs is due to internal  
relativistic shocks propagating in a highly collimated jet, while 
afterglows are produced  via interaction of a forward shock with the 
interstellar medium. In this external shock, the relativistic particles
radiate at all wavelengths, presumably
via the synchrotron mechanism.  While the shock propagates in the
ambient
medium, the expanding plasma progressively decelerates, and becomes
sub-relativistic.
Prompt emission at optical and near-infrared (NIR) wavelengths
simultaneous with the GRB, or delayed by a few seconds, is expected to
take place as a consequence of a reverse shock propagating into the
explosion ejecta. 
This low energy early emission can be very
bright, up to the 5th magnitude in the visual band.  This prediction had
been already reported within a complete theoretical scenario
[14,15,22]
when, on 23 January 1999, an optical flash was  detected
by the ROTSE I robotic telescope starting 22 seconds after a GRB
triggered by both BeppoSAX and BATSE (GRB990123, [3,7]). 
At light maximum the optical transient reached $V
\simeq 9$ [1], implying that the power output in the
optical is $\sim$1\% of that at the high energies. 
The prompt low energy counterpart emission of a GRB must be very closely
related to that of the GRB itself, although possibly also determined by
physical parameters of the ejecta, which instead do not affect the high
energy event.  This may explain the lack of a direct correlation between
gamma-ray and optical fluxes, as inferred by scaling the properties of
many GRBs promptly followed up in the optical to those of GRB990123
[2].\\ 
The field of GRB990123 was rapidly acquired by ROTSE I thanks to the
efficient alert system of the GRB experiment BATSE (GCN), but the prompt
variable optical source simultaneous with GRB990123 had been detected and
identified as the GRB counterpart only after the accurate positioning of
BeppoSAX, which circumscribed the search area within the large ROTSE I CCD
image ($16^{\circ} \times 16^{\circ}$ size). In this detection, the prompt
slew of the ground based small robotic telescope to the GRB position has
played a crucial role: in $\sim$15 minutes, the optical source decayed to $V
\sim 15$, i.e. a factor $\sim$250 in flux, and steadily faded in time.  The
optical and the gamma-ray peaks of the prompt emission were not coincident
in time (Fig. 1), and the decay rate of the optical flash is steeper than
that of the early afterglow (Fig. 2), consistent with 
three different mechanisms, i.e. an internal shock, a reverse shock
propagating into the ejecta and a forward shock propagating in the external
medium, being responsible for the GRB itself, its associated optical flash,
and the early optical afterglow, respectively [16,21].
\begin{figure}
\epsfysize=7cm % fix the y-dimension and scales x-dim. to y-dim.
%\epsfxsize=8cm % fix the x-dimension and scales y-dim. to x-dim.
% Feel free to do the choice you prefer but do not exceed the x-dimension
% of the text lines
\hspace{3.0cm}\epsfbox{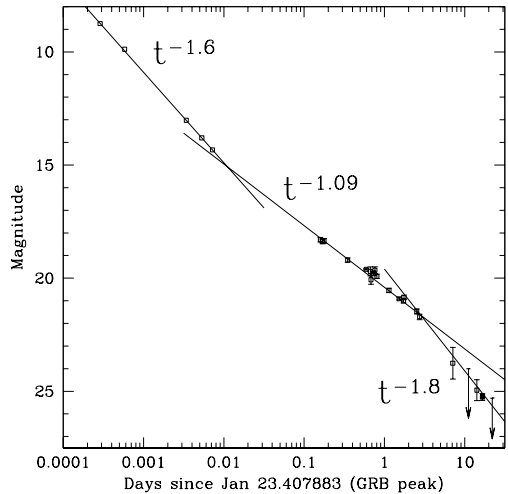} 
%for centering: act on hspace argument 
%\vspace{-0.3cm}
\caption[h]{Light curve in the R band of the optical counterpart of
GRB990123. The early points are the ROTSE I measurements. The points
past 0.1 days are afterglow detections from larger telescopes.
Overplotted onto the data are the power-laws which best fit the
different portions of the light curve (from [11]).}
\end{figure}
The above sequence of actions makes it clear that the chance of
detecting the low-energy counterparts of GRBs are determined by
1) rapid dissemination of accurate GRB error boxes, 2)  prompt
follow-up with an extremely flexible optical/NIR facility.  The former
condition has not been systematically satisfied so far: all past GRB
localizing missions provided either large error boxes after some seconds
delay (BATSE) or arcminute precise error boxes within hours or
days (BeppoSAX, IPN).  The second condition implies reaction times of
tens of seconds, which were not afforded by most robotic facilities
(most of them react in a matter of several minutes).

\section{REM Optical Slitless Spectrograph (ROSS): coupling rapid
spectral evolution of GRBs in the optical and at high energies}

In the near future, three missions for high energy astrophysics will provide
a high number of GRB detections with arcminute precision localizations:
HETE-2, launched in October 2000 (30 to 50 events localized per year);
AGILE, which will become operative in 2002 (about 10 events per year); and
SWIFT, expected for 2003 ($\sim$300 events per year). All of them will
disseminate GRB error boxes centroid coordinates and errors in real time.
This represents a crucial improvement in GRB observations with respect to
both BATSE, which could provide only large ($\sim$50 square degrees)  error
boxes, and BeppoSAX, whose arcminute size error boxes are available no
earlier than 3-4 hours after the GRB.
The real time dissemination of
arcminute error boxes will enable us to begin optical observations
immediately. Since the optical burst spectral distributions may evolve
rapidly, truly simultaneous multi-band observations are required.
Multiplexing the spectral response with rotating photometric filters is not
satisfactory since it would not allow a strictly simultaneous coverage of
the whole optical band.\\
The small size of future GRB error boxes no longer
requires large fields of view.  Therefore, in April 2000 we started a
collaboration with Dr.  Akerlof (Univ. of Michigan) aimed at building
a variant of the last robotic system developed by the ROTSE
team, with
increased focal length and smaller field of view. The inclusion of a narrow
wedge objective prism (replaced, in the current project, by an AMICI prism)  
would have allowed prompt low resolution spectroscopy of
GRBs.  The project was partly funded by the Italian Space Agency (ASI); 
the European Southern Observatory (ESO) approved the installation of our
telescope at La Silla together with two more robotic systems:
TAROT-1S and REM,
devoted to prompt optical and NIR photometry of GRBs, respectively.
In June 2001, the REM team and our team
decided to merge our projects, by accommodating the optical
spectrograph
in the REM telescope.  Following this new design the instrument has been
called REM Optical Slitless Spectrograph (ROSS). A dichroic will allow
simultaneous NIR photometry and low resolution optical spectroscopy. The
larger size of the REM telescope with respect to the original configuration
(60 cm
diameter versus 40 cm) will compensate for light losses due to the dichroic
(about 10\%). REM installation at La Silla is foreseen for the Fall of 2002;
the expected detection rate is of at least 4-6 events per year.
ROSS will measure the optical continuum over the
wavelength range 4500-9000 \AA\ in
$\sim$20 spectral bins $\sim$200 \AA\ wide with a signal-to-noise of
5 or better for a source of average magnitude $V \sim 15$.  The
spectroscopic acquisitions will be sequentially made in logarithmically 
spaced time intervals.  We expect to be able to follow for at least
$\sim$15 minutes
a source of initial brightness up to a factor of 100 lower 
than that of the GRB990123 flash ($V \sim 14$), and fading at the
same rate.\\
The peak of the spectral energy distribution
of a GRB, located in the gamma-ray band during the prompt event (namely from
about 5 to 20 seconds after the explosion), shifts to the optical in a
matter of 2-3 hours, according to the synchrotron shock model [20]. This is
consistent with observations of multiwavelength afterglows of GRBs [10] and
implies that from some minutes to few hours after the GRB, when ROSS
monitoring takes place, the peak energy is close to and eventually crosses
the optical band.  Therefore, at these early epochs the most rapid spectral
changes are expected in the optical, which makes ROSS observations most
critical. As a consequence, we will better understand the onset of the
afterglow and will be able to link its physical characteristics with those
of the prompt event through correlated 
gamma-ray and optical spectral variability. Moreover, we may be able to
discern among emission models, which differ most in their predictions at
early epochs [6,16,20,22].
For bright flashes we should be able to detect time-dependent
obscuration
of optical transients associated with GRBs, predicted to be possibly
present and detectable in the optical at a redshift higher than 1.5.\\
Unlike BeppoSAX, the future gamma-ray missions will be able to
accurately localize also sub-second duration GRBs [13]; fast ROSS
spectroscopy
of these error boxes will therefore effectively address the problem of
the nature of these events.  They may have different types of precursors
and thus different environments and/or redshift distributions than long
GRBs. The fireball mechanism for short bursts may be entirely different
than for events optically identified so far. This would clearly show
up if such bursts were predominantly found far from galactic centers,
this being an expected feature of neutron star merger scenarios. In this
case, optical afterglows would hardly develop, due to the very low
density of the external medium, while optical emission simultaneous to
the GRB would be more easily detectable, making an extremely fast
search crucial. For no detections, we would be able to put useful upper
limits.

\section{Synergy between ROSS and other instruments}

Prompt GRB optical/NIR counterparts carry a physical information as rich
and important as that of the GRB themselves.
The high flux level expected from the very early emission of
these sources makes them a benchmark of the GRB phenomenon;
therefore, it is important to guarantee an effective observational
approach for their study, encompassing photometry and spectroscopy in
the optical and NIR bands.\\
The possible simultaneous operations at La Silla of three instruments for
rapid photometry in the optical (TAROT-1S) and NIR (REM), and rapid
spectroscopy in the optical (ROSS) appear ideal to tackle in a
complete way the investigation on optical/NIR flashes. TAROT-1S and the REM
NIR camera will catch the flash at early times providing transient
coordinates with
sub-arcsec precision.  These can be rapidly disseminated and
utilized for immediate spectroscopy/polarimetry and for further optical
follow-up with a larger telescope. The co-presence of an optical and a
NIR imager ensures the determination of the coordinates even in the case
of dust or Lyman-$\alpha$ absorption which can significantly suppress the
flux in the optical.  
%TAROT-1S will also perform fast photometry. 
ROSS will probe, via rapid, very-low resolution
spectroscopy, the early spectral variability of the optical source
continuum. The scientific goals of the three experiments are complementary
and a logistic cooperation among the facilities will guarantee
efficient communication of fundamental information on the transient
sources (coordinates, flux levels) to the larger ESO telescopes, in case
a follow-up must be activated.\\
ROSS will augment the performance of the UVOT instrument onboard SWIFT in
two crucial ways: it will provide much better coverage of the prompt
optical burst and far better sensitivity to the red portions of the
spectrum. This extended sensitivity is very important for distant events:
at a redshift of 3, the Lyman-$\alpha$ cutoff is approaching the R band.
Such
an event may be invisibly faint in the UV. Although the SWIFT UVOT will
have a selection of filters, only one can be active at a time. Thus, the
simultaneous multichannel photometry and low resolution spectroscopy
afforded by ROSS during the initial optical flash will represent a crucial
complement.\\
There is a very large range of astrophysical observations, unrelated to
GRBs, which could symbiotically take advantage of the REM-ROSS program.
This includes near-Earth objects, variable stars, X-ray transients,
supernovae and AGNs. A fraction of these classes of sources will be also
primary or serendipitous targets of AGILE and SWIFT.

%\vspace{-0.3cm}
\noindent
See the REM Web page for more details on the project and to follow its
future development: http://golem.merate.mi.astro.it/projects/rem/\\

\vspace{-0.3cm}
\noindent
{\bf Note}. 
Since the Vulcano Workshop took place prior to our agreement of
collaboration with the REM project, the present paper refers to a modified
experiment with respect to that described in the talk delivered
during the Workshop. Our original project, now superseded by the ROSS
concept, consisted in an independent robotic telescope, ROTSE IV, and
represented an upgrade of a ROTSE III telescope (see the ROTSE Web page
for more details on ROTSE III, http://www.umich.edu/$\sim$rotse). 

%\begin{figure}
%\vspace{2cm}   % empty space for the figure (2cm in the given example)
%\caption[h]{Type here the caption for the figure.}
%\end{figure}

\vspace{-0.3cm}
\acknowledgements
E. Palazzi would like to thank Franco Giovannelli and the LOC for
the warm hospitality in Vulcano and for organizing a stimulating meeting.

% References. We avoided using the \bibitem commmand since we found it is
% somewhat platform-dependent. We also avoided using the \cite{keyword}
% command since we found it cumbersome. However, if you are an expert 
% LateX user you may use the various LateX tools for the references 
% provided they give the same printout formats of the examples given here.

\end{document}